\documentclass[12pt]{iopart}
\pdfoutput=1

\usepackage{amssymb,amsfonts}	
\usepackage{tikz,tikz-3dplot}\tikzset{>=latex}	
\usepackage{graphicx} 
\usepackage[caption=false,position=top,]{subfig} 
\usepackage{bm}	

\usepackage{hyperref}\hypersetup{colorlinks=true,allcolors=blue}	
\usepackage[all]{hypcap}	

\usepackage{cleveref}	
										
	\crefformat{equation}{equation~#2(#1)#3}
	\crefrangeformat{equation}{equations~#3(#1)#4-#5(#2)#6}
	\crefmultiformat{equation}{equations~#2(#1)#3}{ and~#2(#1)#3}{, #2(#1)#3}{ and~#2(#1)#3}

	\Crefformat{equation}{Equation.~#2(#1)#3}
	\Crefrangeformat{equation}{Equations~#3(#1)#4-#5(#2)#6}
	\Crefmultiformat{equation}{Equations~#2(#1)#3}{ and~#2(#1)#3}{, #2(#1)#3}{ and~#2(#1)#3}
	
	\crefformat{figure}{figure~#2#1#3}
	\crefrangeformat{figure}{figures~#3#1#4-#5#2#6}
	\crefmultiformat{figure}{figures~{#2#1#3}}{ and~#2#1#3}{, #2#1#3}{ and~#2#1#3}
	
	\Crefformat{figure}{Figure~#2#1#3}
	\Crefrangeformat{figure}{Figures~#3#1#4-#5#2#6}
	\Crefmultiformat{figure}{Figures~{#2#1#3}}{ and~#2#1#3}{, #2#1#3}{ and~#2#1#3}
	
	\crefformat{subsection}{subsection~#2#1#3}
	\Crefformat{subsection}{Subsection~#2#1#3}
	
	\crefformat{appendix}{#2#1#3}

\AtBeginDocument{\renewcommand{\ref}[1]{\cref{#1}}}

\begin{document}

\newcommand{\n}[1]{\textbf{#1}}
\renewcommand{\it}[1]{\textit{#1}}
\renewcommand{\d}{\partial}

\title[Energy extraction of a chaotic system in a cyclic process]{Energy extraction of a chaotic system in a cyclic process: a Szil\'ard Engine perspective}

\author{Artur Soriani$^1$ and Marcus V. S. Bonan\c ca$^2$}

\address{Instituto de F\'{\i}sica `Gleb Wataghin', Universidade Estadual de Campinas, 13083-859, Campinas, S\~ao Paulo, Brazil}

\ead{
\begin{tabular}{l}
$^1$ a164189@dac.unicamp.br\\
$^2$ mbonanca@ifi.unicamp.br
\end{tabular}
}

\date{\today}

\begin{abstract}
Inspired by the available examples of Microcanonical Szil\'ard Engines and by the original Szil\'ard Engine, we devise a system with two degrees of freedom whose ensemble average energy, starting with a microcanical ensemble, decreases after a cyclic variation of its external parameters.
We use the Ergodic Adiabatic Theorem to motivate our cycle and numerical simulations to check the decrement in the average energy.
We then compare our system to the aforementioned Szil\'ard Engines, Microcanonical or not, and speculate about symmetry breaking being the cause of energy extraction in  cyclic processes, even when non-integrability and chaos are present.
\end{abstract}


\maketitle


\section{Introduction}\label{sec:intro}

In 1871, James Clerk Maxwell described a simple thermodynamic system that, by the action of an intelligent being with access to the microscopic state of the system, would violate the Second Law of Thermodynamics, as it was understood \cite{maxwell}.
This intelligent being, later named Maxwell's Demon, seemed like an odd supposition, as microscopic information is never available to us.
Later, in 1929, Leo Szil\'ard proposed a cleaner version of Maxwell's Demon, one where the demon is simplified to an external agent that, based on a measurement made on a system in contact with a single thermal reservoir, exerts a cyclic thermodynamic process on this system and extracts energy from it, going against the Kelvin-Planck statement of the second law.
Szil\'ard's engine \cite{szilard}, as Szil\'ard's thought experiment became known, gives us a more quantifiable version of Maxwell's Demon and one where the importance of information is fully flashed out.

Szil\'ard's original engine is simple, but ingenious.
Consider a one particle gas inside a box connected to a heat reservoir, ensuring that processes the gas goes through are isothermal.
At some point, a barrier is inserted in the middle of the box so as to divide it in two parts and the particle gets trapped in one of the sides.
The insertion of the barrier, in principle, has no energy cost attached to it, or at least the energy cost can be made arbitrarily small.
A measure is then made to determine which side of the box the particle got trapped in.
Knowing which side the particle is, we now let the barrier act as a piston and, by moving the barrier, the one particle gas expands or compresses, meanwhile the particle collides with the barrier and exerts pressure.
More specifically, we let the gas expand quasi-statically to its full original volume and, while colliding with the barrier, the particle exerts work and loses energy.
When the barrier reaches the end of the box, it is removed (also with no energy cost, like the barrier insertion) and the cycle is finished.
Kelvin-Planck's statement of the second law says that ``\it{it is impossible to devise a cyclically operating device, the sole effect of which is to absorb energy in the form of heat from a single thermal reservoir and to deliver an equivalent amount of work}''.
But that is exactly what Szil\'ard's engine does: the whole cycle operates at constant temperature and the energy the particle loses can be extracted and stored (as potential energy of a weight, for example).

Later contributions by Landauer \cite{landauer} and Bennett \cite{bennett} led to the conclusion that the Second Law of Thermodynamics is still valid once the energetics of information storage and information erasure are considered.
Finally, Sagawa and Ueda generalized these ideas \cite{sagawa_ueda1,sagawa_ueda2,sagawa_ueda3,sagawa_ueda4,sagawa_doc}.
They formulated the \it{Second Law of Information Thermodynamics}, where thermodynamic quantities and information quantities are treated on equal footing.
In turn, the work by Sagawa and Ueda unfolded a myriad of other works \cite{price_etal,cao_feito,bannerman_etal,jacobs,ponmurugan,horowitz_vaikun,abreu_seifert} about the Thermodynamics of feedback processes (processes dependent on a previous measurement, like the original Szil\'ard Engine), and, in particular, we now have autonomous Maxwell demons \cite{mandal_jarzynski,barato_seifert1,deffner_jarzynski,barato_seifert2,lu_etal}, examples where the demon is itself a physical sub-system, part of a bigger system.

In recent years, there has been a resurgence of interest in the implementation of different versions of Maxwell Demons, i.e. systems that, through use of measurement and information, defy certain statements of the Second Law of Thermodynamics.
A quantum formulation of Szil\'ard's Engine \cite{quantum_szilard_engine} has only recently been developed.
Experimental verifications of the relation between information and energy have also been made recently \cite{universal,exp1,exp2,exp3}.
It has even been shown, based on fluctuation theorems, that feedback processes constitute a kind of symmetry breaking on the system's phase-space and that this symmetry breaking is ultimately the mechanism that allows energy extraction from a single heat reservoir to happen \cite{universal,revisited,phase}, at least for processes carried out at constant temperature.

More important to the scope of this work are Maxwell Demons that start with microcanonicaly sampled initial conditions, instead of the isothermal condition always present in the original Szil\'ard Engine.
For instance, Sato \cite{sato} provided an example of a one-dimensional system whose ensemble average energy, starting with a microcanonical ensemble, decreases after an operation.
Later, Microcanonical Szil\'ard Engines were proposed \cite{cooling,modeling,bonanca_acconcia}, versions of the Szil\'ard Engine where it is possible to extract energy from a single heat reservoir in a cycle by using the information acquired from a measurement of the energy of the system (not position, like in the original Szil\'ard's Engine) in every realization of the cycle (not simply on average, like in Sato's example).
It has yet to be proven whether or not symmetry breaking is responsible for energy extractions in Microcanonical Szil\'ard Engines.
If it is, we expect more complex examples of this kind of engine to exist.
For a quantum example of work extraction from a microcanonical bath in the thermodynamic limit, see reference \cite{allahverdyan_hovhannisyan}.

This work is laid out as follows: in \ref{sec:therm_cycle}, we describe the general thermodynamic cycle we use; in \ref{sec:working_substance}, we present the specific Hamiltonian we work with; in \ref{sec:mechanical_cycle}, we present the mechanical cycle described by the cyclic variation of the external parameters of said Hamiltonian; in \ref{sec:cycle_simulations_results}, we apply the Ergodic Adiabatic Theorem and numerical simulations to check the energy variations within the mechanical cycle; in \ref{sec:discussion}, we compare our model with other models and discuss its achievements; in \ref{sec:conclusion}, we give our concluding remarks.

\section{The thermodynamic cycle}\label{sec:therm_cycle}

In this work, we aim to provide a system with two degrees of freedom whose ensemble average energy, starting with a microcanical ensemble, decreases after cyclic variation of its external parameters.
Two degrees of freedom may not seem as much of an improvement from one, but it already introduces more complex dynamic concepts like non-integrability and chaos, which are present in most thermodynamic systems.
Consider the following thermodynamic cycle:

\begin{enumerate}

\item[1.] We start with an equilibrium ensemble of our system with Hamiltonian $H$ in contact with a heat reservoir at temperature $T$.

\item[2.] We disconnect each element of the ensemble from the reservoir.
At this point, the ensemble in question is a canonical ensemble at temperature $T$.

\item[3.] We measure the energy of each element in our ensemble and sort them according to the value $E$ of energy measured.
The result is then a set of sub-ensembles of well-defined energy, i.e., microcanonical sub-ensembles.

\item[4.] We implement a mechanical cycle in a given sub-ensemble \emph{only if} its average energy \emph{decreases} after it.
By mechanical cycle, we mean a cyclic variation of external parameters $\bm{\lambda}$ in the system's Hamiltonian.
The final result is a decrease of the average energy of the total ensemble since we have either decreased or left constant the average energies of the sub-ensembles.

\item[5.] We reconnect the ensemble with the heat reservoir and let it return to equilibrium.

\end{enumerate}

\begin{figure*}

\newcommand{\scale}{1}

\centering

\subfloat[]{
\begin{tikzpicture}[scale = \scale, every node/.style={transform shape}]

\draw (-4.5,1.5) rectangle (1,-1.5);

\draw[fill=red!20] (-3,0) circle (1cm);
\node at (-3,0) {$T$};

\draw[thick] (-2,0) -- (-1,0);

\draw (-1,0) -- (-1,.5) -- (0,.5) -- (0,-.5) -- (-1,-.5) -- cycle;
\node at (-.5,0) {S};

\end{tikzpicture}
}
\raisebox{-1.8cm}{
\begin{tikzpicture}[scale = \scale, every node/.style={transform shape}]
\node {\large $\bm{\Rightarrow}$};
\end{tikzpicture}
}
\subfloat[]{
\begin{tikzpicture}[scale = \scale, every node/.style={transform shape}]

\draw (-4.5,1.5) rectangle (1,-1.5);

\draw[fill=red!20] (-3,0) circle (1cm);
\node at (-3,0) {$T$};

\draw[thick] (-2,0) -- (-1.6,0);
\draw[thick] (-1.4,0) -- (-1,0);
\draw[red,thick] (-1.7,-.1) -- (-1.5,.1);
\draw[red,thick] (-1.5,-.1) -- (-1.3,.1);

\draw (-1,0) -- (-1,.5) -- (0,.5) -- (0,-.5) -- (-1,-.5) -- cycle;
\node at (-.5,0) {S};

\end{tikzpicture}
}

\bigskip

{\large \bm{$\Uparrow$} \hspace{6.3cm} \bm{$\Downarrow$}}

\captionsetup[subfigure]{position=b}

\renewcommand{\thesubfigure}{d}

\subfloat[]{
\begin{tikzpicture}[scale = \scale, every node/.style={transform shape}]

\draw (-4.5,1.5) rectangle (1,-1.5);

\node at (-3,.5) {Cyclic};
\node at (-3,0) {variation};
\node at (-3,-.5) {of $\bm{\lambda}(t)$};

\draw (-1,0) -- (-1,.5) -- (0,.5) -- (0,-.5) -- (-1,-.5) -- cycle;
\node at (-.5,0) {S};

\end{tikzpicture}
}
\raisebox{1.35cm}{
\large
\bm{$\Leftarrow$}
}
\renewcommand{\thesubfigure}{c}
\subfloat[]{
\begin{tikzpicture}[scale = \scale, every node/.style={transform shape}]

\draw (-4.5,1.5) rectangle (1,-1.5);

\node at (-3,.25) {Measurement of};
\node at (-3,-.25) {the energy of S};

\draw (-1,0) -- (-1,.5) -- (0,.5) -- (0,-.5) -- (-1,-.5) -- cycle;
\node at (-.5,0) {S};

\end{tikzpicture}
}

\caption{
(Color online) Schematic representation of the thermodynamic cycle proposed.
We start in the top left, panel (a), with each copy $S$ of the system of our ensemble connected to heat reservoir at temperature $T$.
We then disconnect $S$ from the reservoir, on panel (b).
Each element of our ensemble now occurs with probability given by the Boltzmann weight.
Moving on to panel (c), we measure the energy of each element in the ensemble and organize them in microcanonical sub-ensembles of well defined energy.
Given our mechanical cycle described by $\bm{\lambda}(t)$, on panel (d) we submit it only on elements of sub-ensembles that have their average energy decreased after the mechanical cycle.
Finally, we reconnect the whole initial ensemble with the heat reservoir, returning it to canonical equilibrium.
This brings us back to panel (a) of the figure and finishes the thermodynamic cycle.
}

\label{fig:thermodynamic_cycle}

\end{figure*}
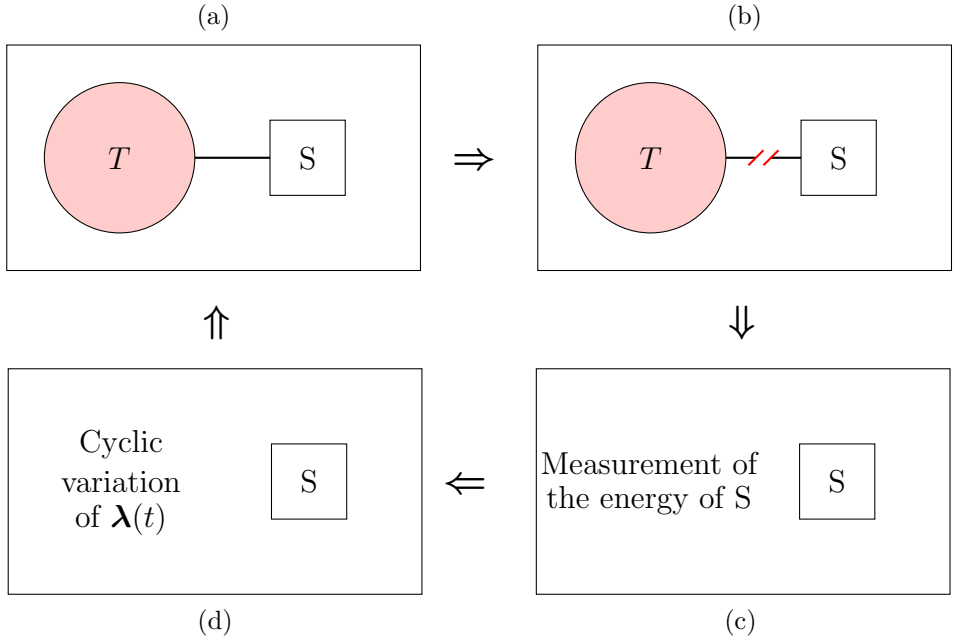

This thermodynamic cycle in schematized in \ref{fig:thermodynamic_cycle}.
It is obviously a cycle and the energy that the sub-ensemble loses during the cycle is extracted as work.
We are transforming the energy absorbed of a single heat reservoir into work in a cycle, in clear contradiction to the Kelvin-Planck statement of the second law.

\section{Working substance with two degrees of freedom}\label{sec:working_substance}

Our working substance will be given by the following QS Hamiltonian \cite{carnegie_percival}
\begin{equation}\label{eq:QS}
H^{\mbox{\scriptsize QS}}(\bm{z},a) = \frac{p_x^2}{2} + \frac{p_y^2}{2} + \frac{a}{4}\left( x^4 + y^4 \right) + \frac{x^2 y^2}{2},
\end{equation}
where QS stands for \it{quartic system}, $\bm{z} = (x,y,p_x,p_y)$ is a point in four-dimensional phase-space and $a$ is an always positive external parameter responsible for determining the form of the potential well.
When $a = 1$, the last two terms of \ref{eq:QS} can be combined in a single term $\frac{(x^2 + y^2)^2}{4}$ and we have a central potential, which in turn means that angular momentum is conserved, serving as a second constant of motion (the first is the energy) and making the system integrable.
On the other hand, when $0 < a < 1$, the system has only one constant of motion and is not integrable.
In addition to that, for low values of $a$, like $a = 0.1$, the system can be considered ergodic for all practical purposes.
Perhaps even more interestingly, a generic ensemble of this system with $a = 0.1$ has the property of auto-relaxation: when allowed to evolve by itself, it will relax to a microcanonical ensemble after time $\tau_R$, called the relaxation time.

\begin{figure*}

\centering

\subfloat[\label{fig:poincare_sections_1.0}]{
\includegraphics[scale=.65]{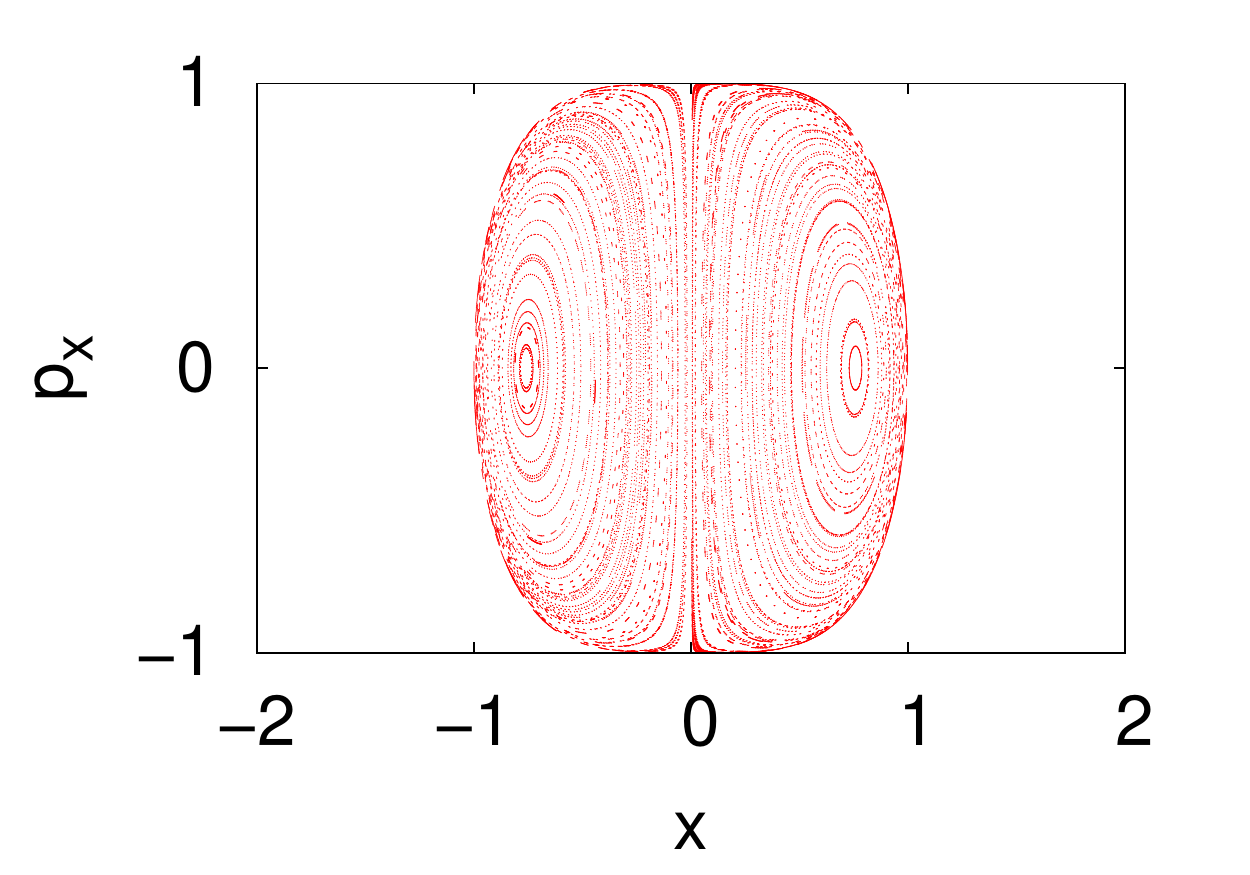}
}

\subfloat[\label{fig:poincare_sections_0.5}]{
\includegraphics[scale=.65]{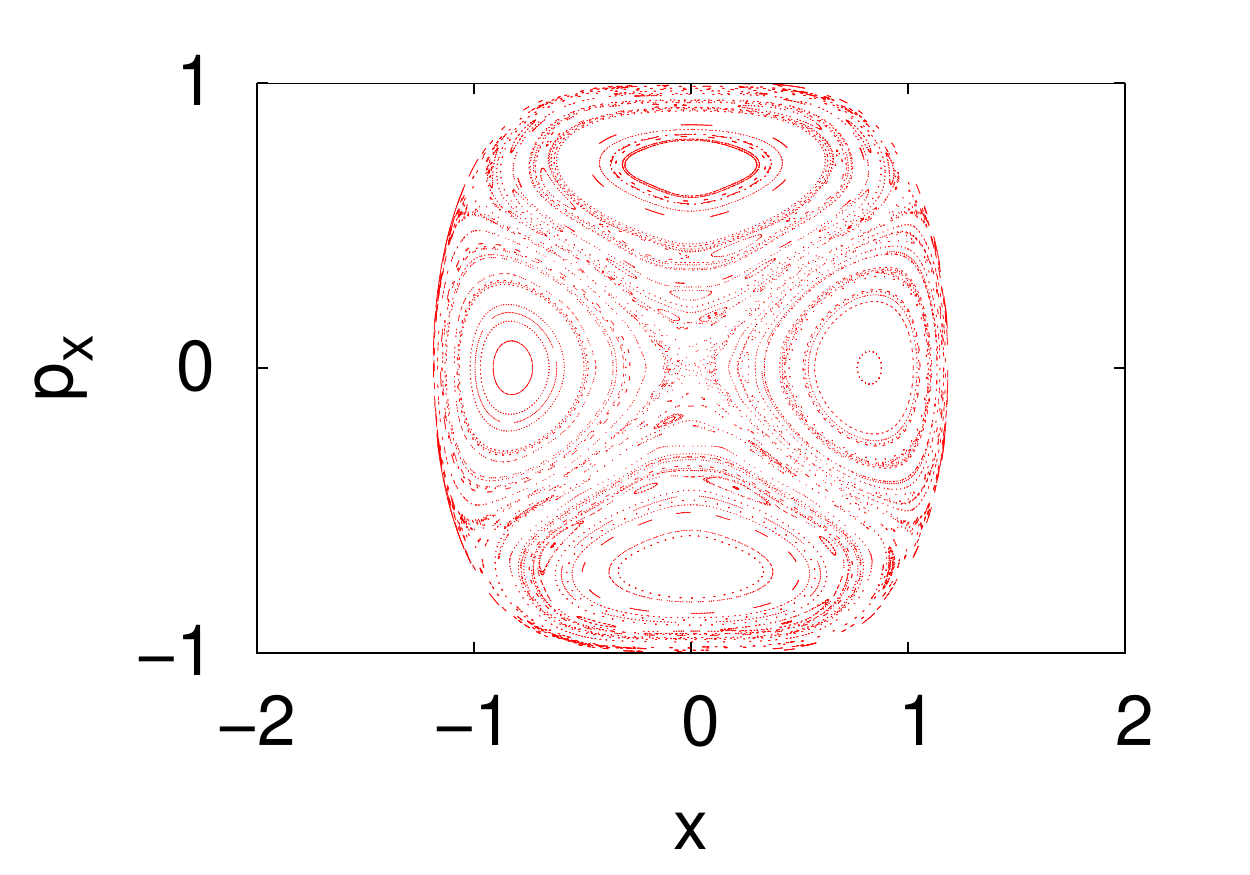}
}

\subfloat[\label{fig:poincare_sections_0.1}]{
\includegraphics[scale=.65]{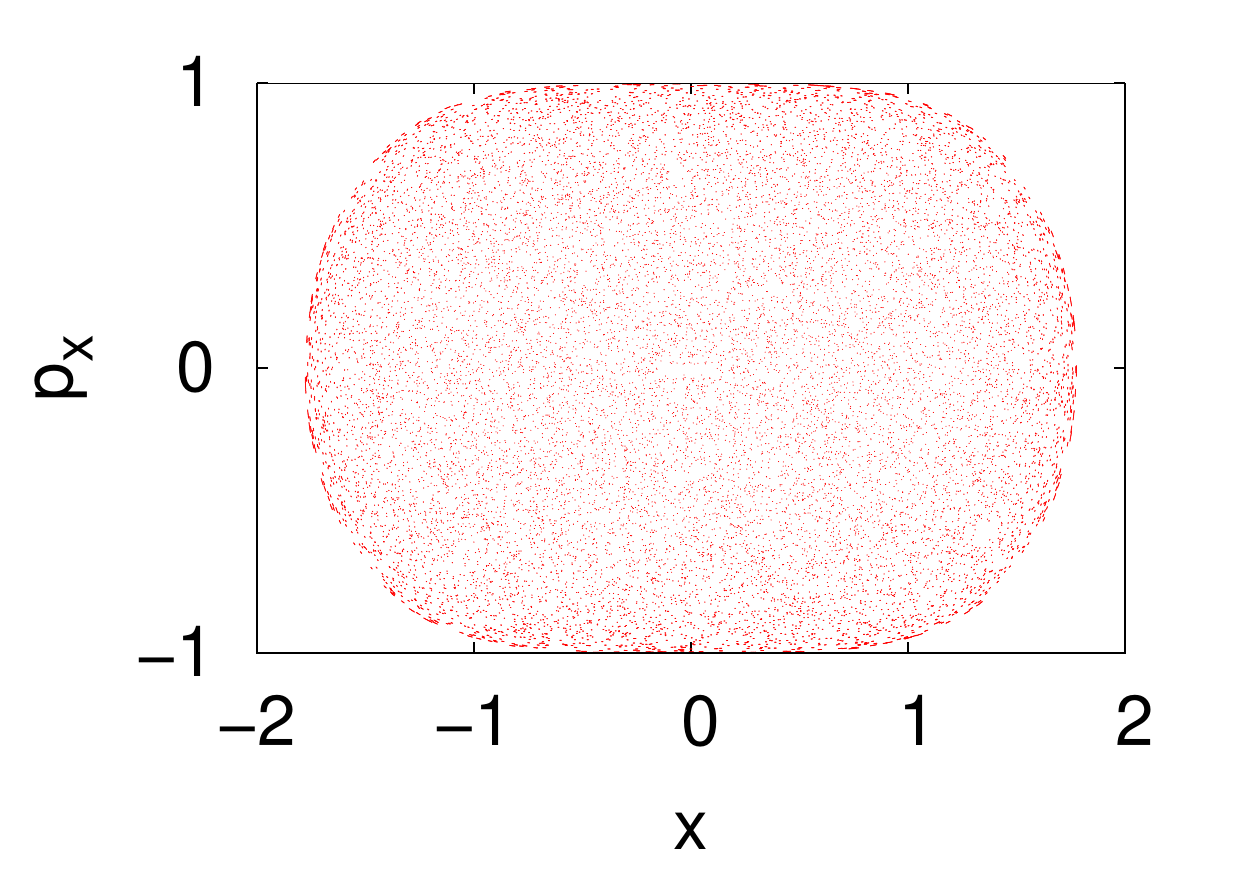}
}

\caption{
(Color online) Poincar\'e sections for the QS Hamiltonian of \ref{eq:QS} in the plane $y = 0$ for $p_y > 0$ with $E = 0.5$ for (a) $a = 1$; (b) $a = 0.5$; (c) $a = 0.1$.
These sections were obtained with the use of a symplectic numerical integrator \cite{forest}, with $100$ initial conditions and an elapsed time $\tau = 1000$.
\Cref{fig:poincare_sections_1.0,fig:poincare_sections_0.5} do not respect the reflection symmetry in the $p_x$ axis, present in \ref{eq:QS}, because we did not use symmetrical initial conditions.
}

\label{fig:poincare_sections}

\end{figure*}

Poincar\'e sections of the system for different values of $a$ are shown in \ref{fig:poincare_sections}.
These are reduced two-dimensional phase spaces of the system that maintain all the characteristics that the original four-dimensional phase space has \cite{lichtenberg}.
For $a = 1$ (\ref{fig:poincare_sections_1.0}), all trajectories display complete regular behavior, as all of them are periodic and have constant angular momentum.
There is a clear division between trajectories with positive angular momentum (in this case, with $x > 0$) and trajectories with negative angular momentum ($x < 0$).
For $a = 0.5$ (\ref{fig:poincare_sections_0.5}), we have a mixture of regular and irregular behavior: some trajectories are periodic, as can be seen in the four lobes with circular shapes, and some are not periodic, concentrated in the middle of the figure.
For $a = 0.1$ (\ref{fig:poincare_sections_0.1}), all the trajectories display approximately ergodic behavior, as all of the phase space points are well distributed over the the entire surface.

\begin{figure*}

\centering

\includegraphics[scale=1.2]{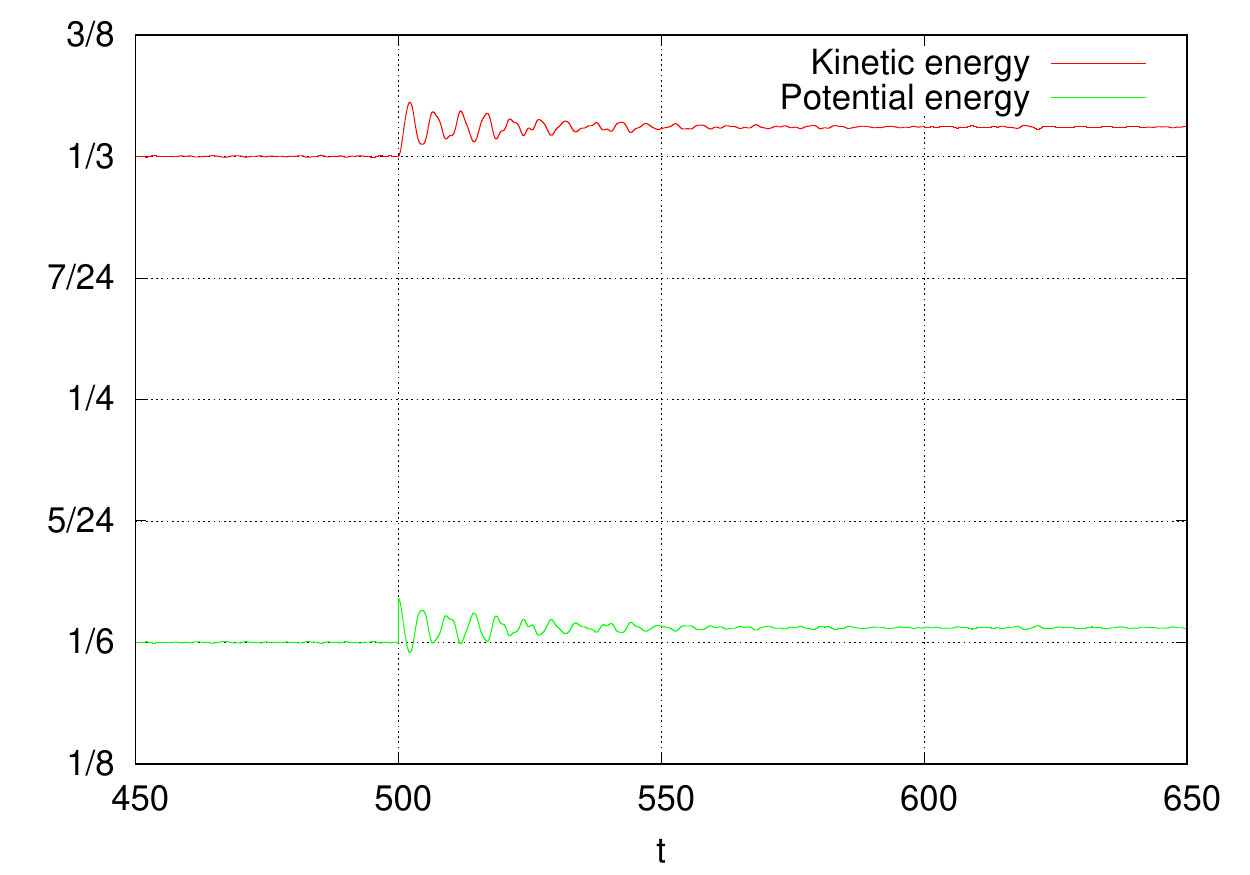}

\caption{
(Color online) Ensemble averages of kinetic and potential energies, the first two terms and the last two terms in \ref{eq:QS}, respectively.
We sampled $10^6$ initial conditions of the QS Hamiltonian with $E = 0.5$ and $a = 0.1$.
We evolved this microcanonical ensemble with a sympletic numerical integrator \cite{forest} for $500$ units of time, suddenly changed $a$ from $0.1$ to $0.12$ and then let the system evolve for $500$ more units of time.
The system relaxes to a micrononical state after less than 100 units of simulation time.
}

\label{fig:SBR}

\end{figure*}

To demonstrate the property of auto-relaxation, we sampled a microcanonical ensemble of the Hamiltonian of \ref{eq:QS} with initial energy $E = 0.5$ for $a = 0.1$ and changed it abruptly to $a = 0.12$.
The average kinetic and potential energies of this sampling are displayed in \ref{fig:SBR}.
In it, we can see that the ensemble relaxes to a microcanonical ensemble (with stationary averages in agreement with analytical calculations in the microcanonical ensemble, see \ref{sec:averages_microcanonical_ensemble}) again in less that 100 units of simulation time, so we can make $\tau_R = 100$.
Since the dynamics of the Hamiltonian of \ref{eq:QS} is scalable with energy (see equation (18) of reference \cite{bonanca}), so are its time scales and we can conclude that the relaxation time for other values of energy is $\tau_R(E') = \left(\frac{0.5}{E'}\right)^{1/4} 100$. 
This relaxation time serves as characteristic time scale of the Hamiltonian and any other time scale should be compared to it.

We want to make use of the separation between trajectories with positive and negative angular momentum, as shown in \ref{fig:poincare_sections_1.0}, in order to extract energy from the system.
However, cyclic variation of the parameter $a$ in \ref{eq:QS} is not enough, as Fermi acceleartion phenomena with only one external parameter show \cite{robust}.
To this end, we add a term to \ref{eq:QS} proportional to the angular momentum $L = x p_y - y p_x$,
\begin{equation}\label{eq:2DMSE}
H(\bm{z},a,b) = \frac{p_x^2}{2} + \frac{p_y^2}{2} + \frac{a}{4} \left( x^4 + y^4 \right) + \frac{x^2 y^2}{2} + b\left( x p_y - y p_x \right),
\end{equation}
and manipulate $a$ and $b$ as we see fit.
This newly added term $b\left( x p_y - y p_x \right)$, for any value of the external parameter $b$, does not interfere with the angular momentum conservation for $a = 1$.

\section{The mechanical cycle}\label{sec:mechanical_cycle}

A mechanical cycle is implemented in this system by attributing temporal dependence to $a(t)$ and $b(t)$ during a time interval $\tau$ and making sure that $a(\tau) = a(0)$ and $b(\tau) = b(0)$.
We have developed a feedback cycle that depends on a measurement of angular momentum of a given trajectory.
The mechanical cycle we will use, divided in three steps, is as follows.

\begin{enumerate}

\item Starting with $a = 0.1$ and $b = 0$, in an ergodic regime, $a$ increases linearly with time until it reaches $a = 1$ and $b$ does not change, between times $t = 0$ and $t = \tau/4$;

\item A measurement of the sign of the angular momentum $L$ of the trajectory is made. Then, between times $t = \tau/4$ and $t = \tau/2$, $a$ does not change and, if $L > 0$, $b$ decreases linearly with time until it reaches $b = -b_M$, else if $L < 0$, $b$ increases linearly with time until it reaches $b = b_M$, with $b_M > 0$;

\item And finally, between $t = \tau/2$ and $t = \tau$, $a$ decreases linearly with time from $a = 1$ to $a = 0.1$ and simultaneously $b$ returns to its original value, $0$, either from $b_M$ or $-b_M$, finishing the mechanical cycle.

\end{enumerate}

Or, more succinctly,

\[
a(t) = \left\{
\begin{array}{ccccccc}
0.1 + 0.9 \frac{4t}{\tau},	&  \mbox{if} &    0   & < & t & < & \tau/4;\\
1,							&  \mbox{if} & \tau/4 & < & t & < & \tau/2;\\
1.9 - 0.9 \frac{2t}{\tau},	&  \mbox{if} & \tau/2 & < & t & < & \tau;
\end{array}
\right.
\]
and
\[
b(t) = \left\{
\begin{array}{ccccccc}
0,										&  \mbox{if} &    0   & < & t & < & \tau/4;\\
b_M \left( -1 + \frac{4t}{\tau} \right),	&  \mbox{if} & \tau/4 & < & t & < & \tau/2;\\
b_M \left(  2 - \frac{2t}{\tau} \right),	&  \mbox{if} & \tau/2 & < & t & < & \tau.
\end{array}
\right.
\]

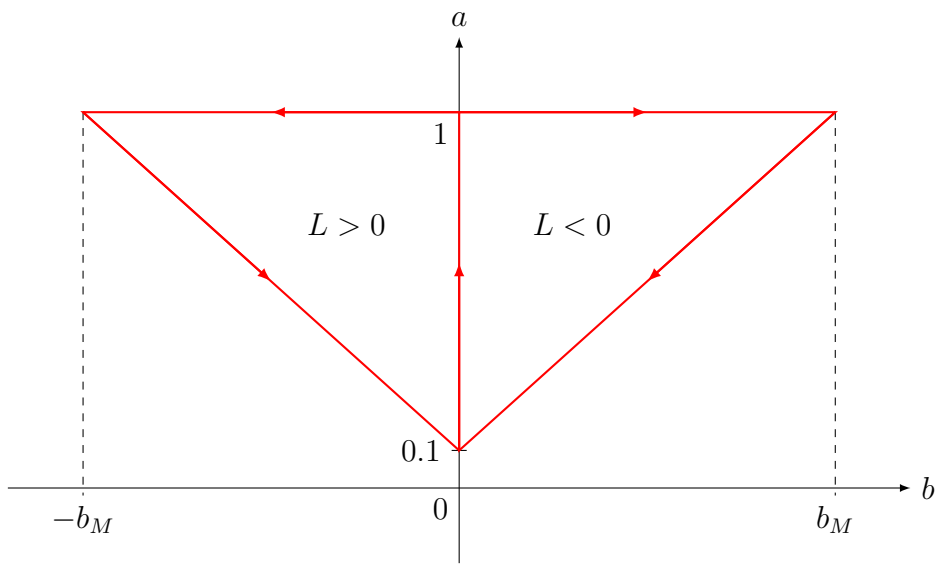
\begin{figure*}

\centering

\begin{tikzpicture}[scale = 5]

\draw[->] (-1.2,0) -- (1.2,0) node[anchor = west]{$b$};
\draw[->] (0,-.2) -- (0,1.2) node[anchor = south]{$a$};

\node[anchor = north east] at (0,0) {$0$};
\draw (.02,.1) -- (-.02,.1) node[anchor = east]{$0.1$};
\node[anchor = north east] at (0,1) {$1$};
\draw[dashed] (1,1) -- (1,-.02) node[anchor = north]{$b_M$};
\draw[dashed] (-1,1) -- (-1,-.02) node[anchor = north]{$-b_M$};

\draw[red, thick] (0,.1) -- (0,1) -- (1,1) -- cycle;
\draw[red, thick] (0,1) -- (-1,1) -- (0,.1);

\draw[->, red, thick] (0,.1) -- (0,.6);
\draw[->, red, thick] (0,1) -- (.5,1);
\draw[->, red, thick] (0,1) -- (-.5,1);
\draw[->, red, thick] (1,1) -- (.5,.55);
\draw[->, red, thick] (-1,1) -- (-.5,.55);

\node at (.3,.7) {$L < 0$};
\node at (-.3,.7) {$L > 0$};

\end{tikzpicture}

\caption{
(Color online) Graphic representation of the protocols $a(t)$ and $b(t)$ to be implemented in the system of Hamiltonian given by \ref{eq:2DMSE}.
After a measurement of the angular momentum $L$ of the trajectory when $a = 1$ and $b = 0$, we have two possible protocols: the one on the left for $L > 0$ and the one on the right for $L < 0$.
}

\label{fig:cycle}

\end{figure*}

A symmetry is clearly achieved and broken during this protocol: angular momentum conservation.
This symmetry can be easily visualized in \ref{fig:poincare_sections_1.0}, where trajectories of positive and negative angular momentum are well divided.
Hence, the first step of this cycle, responsible for splitting the initial sub-ensemble in two regions of well defined sign of angular momentum, is analogous to the barrier insertion in the Szil\'ard engine.
The attainment and breaking of symmetry in our mechanical cycle is intentional and will be discussed in \ref{sec:discussion}.
The most crucial step of our mechanical cycle is step 2, and it is the reason why we added the angular momentum term to \ref{eq:2DMSE}.
Step 2 is responsible for diminishing the ensemble's mean energy, and later in this section we will show exactly how.

We implement this mechanical cycle quasi-statically, which means $\tau \to \infty$ (or, realistically speaking, $\tau \gg \tau_R$), and make use of the Ergodic Adiabatic Theorem as an attempt to trace the energy variation of a trajectory in this mechanical cycle.
The Ergodic Adiabatic Theorem states that the phase space volume $\Omega(E,\bm{\lambda})$ of a surface of constant energy $H(\bm{z},\bm{\lambda}) = E$ ($\bm{\lambda}$ being a vector containing all the external parameters considered), given by
\begin{equation}\label{Omega}
\Omega(E,\bm{\lambda}) = \int \Theta\bm{(} E - H(\bm{z},\bm{\lambda}) \bm{)} d\bm{z},
\end{equation}
where $\Theta$ is the Heaviside Theta function ($\theta(x) = 1$ if $x \geq 0$ and $\theta(x) = 0$ otherwise), is conserved during a process, as long as the process is quasi-static and the system remains ergodic throughout the entire process \cite{ergodic_adiabatic,ott}.
However, our system does not meet the criteria for this theorem, as we know that our system is integrable for $a = 1$ and therefore not ergodic, but, as we will show, the theorem still gives us a decent prediction for the energy values, at least on average.
In our case, $\bm{\lambda} = (a,b)$ and, even though \ref{Omega} cannot be calculated explicitly for all values of $a$ and $b$, it can be in a few specific cases.
For example,
\begin{equation}\label{eq:Omega_a}
\Omega(E,a,0) = \frac{16\pi}{3} \sqrt{\frac{2}{1-a}} E^{3/2} F\left( \sin^{-1}\sqrt{\frac{1-a}{1+a}} \left| \frac{1+a}{1-a} \right.\right),
\end{equation}
where $F\left( \phi_0 \left| k^2 \right. \right) = \int_0^{\phi_0} \frac{d\phi}{\left( 1 - k^2\sin^2\phi \right)^{1/2}}$ is the incomplete elliptic integral of the first kind, and
\begin{equation}\label{eq:Omega_b}
\Omega(E,1,b) = \frac{\pi^2}{3}\left( \left( b^4 + 4E \right)^{3/2} + b^2 \left( b^4 + 6E \right) \right).
\end{equation}
Both results agree when $a = 1$ and $b = 0$, $\Omega(E,1,0) = \frac{8\pi^2}{3} E^{3/2}$.

\section{Mechanical cycle simulations and results}\label{sec:cycle_simulations_results}

This sections is divided into two subsections: \ref{subsec:each_step}, where we present the energy variation in each step of the mechanical cycle, and \ref{subsec:entire_cycle}, where we present the energy variation in the entire mechanical cycle.

\subsection{Energy variation in each step of the mechanical cycle}\label{subsec:each_step}

\subsubsection{First step:}

During the first step of the mechanical cycle, the system goes from an ergodic state to an integrable one.
Denote by $E_1$ the energy of a trajectory at the beginning of the first step and by $E_2$ the energy of this same trajectory at the end of the first step.
If adiabatic invariance holds true to the quasi-static evolution of this protocol, we know that
\[
\Omega(E_1,0.1,0) = \Omega(E_2,1,0).
\]

Using \ref{eq:Omega_a} and solving for $E_2$,
\begin{equation}\label{eq:e2}
E_2 = \left( \frac{2f(0.1)}{\pi} \right)^{2/3} E_1,
\end{equation}
where
\[
f(a) = \sqrt{\frac{2}{1-a}} F\left( \sin^{-1}\sqrt{\frac{1-a}{1+a}} \left| \frac{1+a}{1-a} \right.\right).
\]

\begin{figure*}

\centering

\includegraphics[scale=.7]{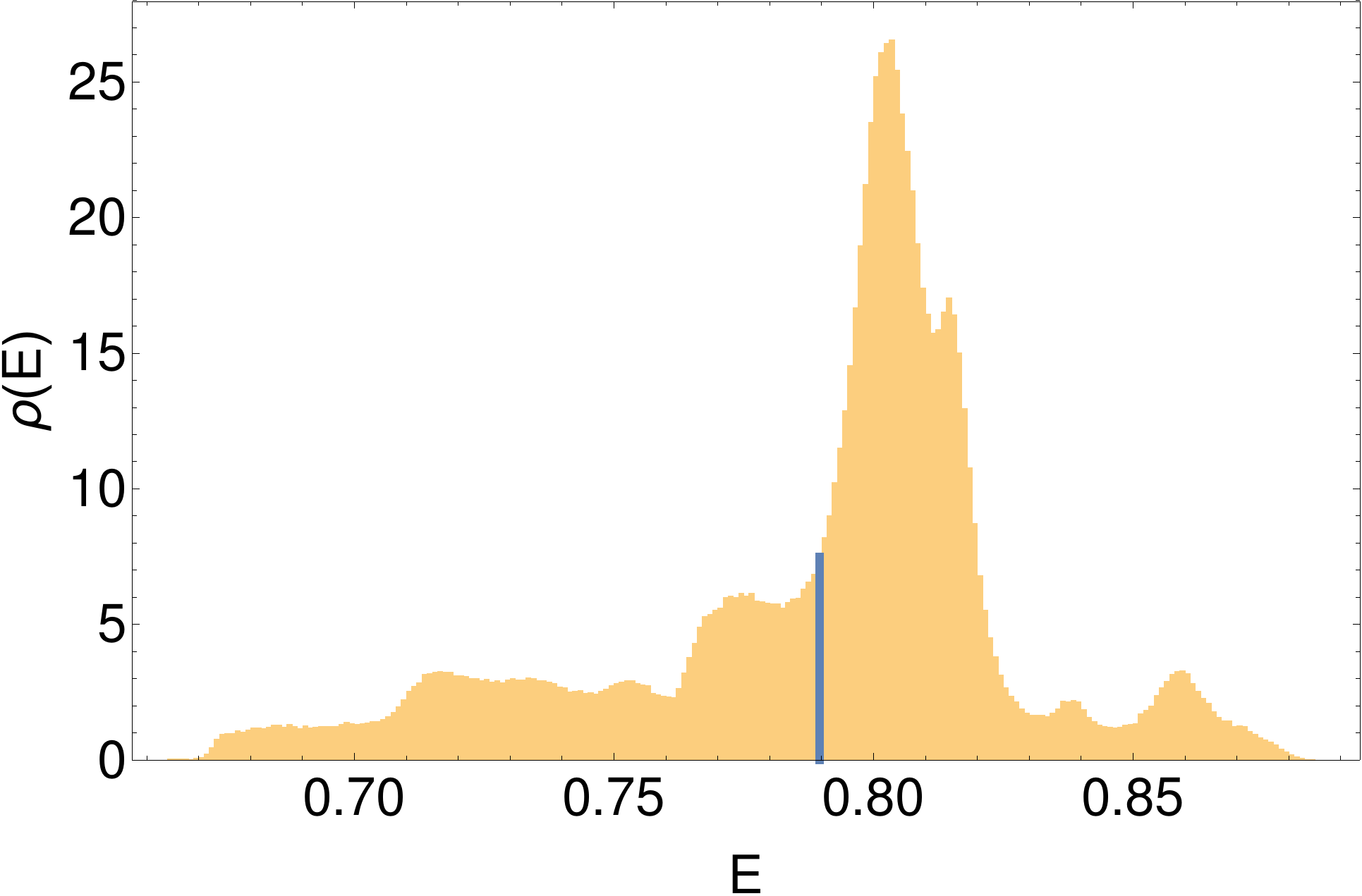}

\caption{
(Color online) Histogram of energies $E$ after the first step of the mechanical cycle, where $\rho(E)$ is the energy distribution obtained from numerical simulations.
We sampled $10^6$ microcanonical initial conditions with energy $E = 0.5$ and evolved them by the first step of the protocol using the sympletic integrator of reference \cite{forest} in a simulation time of $\tau = 10^3 \tau_R$.
Simulations with different switching times $\tau$ were conducted, but the general outline of the histograms are always the same.
The blue line represents the expected energy value from the Ergodic Adiabatic Theorem.
The average energy of the system after the evolution is 0.7896, in excellent agreement with the expected value 0.7870 from equation \ref{eq:e2}.
Of course, the exact energy value of each trajectory after the evolution may not be close to the expected value, and that may be because this process is not perfectly adiabatic, as that would require $\tau \to \infty$.
}

\label{fig:hist_e2}

\end{figure*}

It must be reiterated, however, that we can give no assurance to the validity of \ref{eq:e2}, as the Ergodic Adiabatic Theorem requires ergodicity to be on effect at all times of the evolution, and that is not the case here.
We can, however, test this equation.
Using a symplectic numerical integrator \cite{forest}, we sampled $10^6$ initial conditions with energy $E_1 = 0.5$ and evolved them in time with $b = 0$ and $a$ increasing linearly from $0.1$ to $1$ in an elapsed time $\tau = 10^3 \tau_R$ (as we, of course, cannot achieve true quasi-staticity $\tau \to \infty$ in numerical simulations, we have to at least make sure that the switching time $\tau$ is much greater than a natural time scale of the system, and the relaxation time $\tau_R$ is a good contender for such a time scale).
\Cref{fig:hist_e2} shows a histogram of energies of the aforementioned simulation.
Even though the energies of the trajectories are not the same as the expected value, their average value (0.7896) agrees very well with the expected value (0.7870) from \ref{eq:e2}.

\subsubsection{Second step:}

During the second step of the protocol, $a = 1$ at all times, while $b$ increases (decreases) from $0$ to $b_M$ ($-b_M$).
This means that the system is integrable at all times, and so we cannot use the Ergodic Adiabatic Theorem.
However, there is a much simpler way to predict how the energy will change in this step.

Suppose a Hamiltonian can be written as
\[
H(\bm{z},\bm{\lambda}) = H_0(\bm{z}) + g(\alpha,\bm{\lambda}),
\]
where $g$ is a generic differentiable function, $\alpha = \alpha(\bm{z})$ is a constant of motion for the Hamiltonian $H_0$ (and, by extension, a constant of motion for $H$) and time dependence may be introduced through $\bm{\lambda} = \bm{\lambda}(t)$, the vector of external parameters.
The energy difference of a trajectory, between times $t_1$ and $t_2$, is
\[
\Delta E = \int_{t_1}^{t_2} \frac{d H(\bm{z},\bm{\lambda})}{dt} dt.
\]
We know that $\frac{dH}{dt} = \frac{\d H}{\d t}$, so we can write
\[
\Delta E = \int_{t_1}^{t_2} \frac{\d H(\bm{z},\bm{\lambda})}{\d t} dt = \int_{t_1}^{t_2} \bm{\nabla}_{\lambda} g(\alpha,\bm{\lambda}) \cdot \frac{d\bm{\lambda}}{dt} dt,
\]
where $\bm{\nabla}_{\lambda}$ is the gradient operation with respect to $\bm{\lambda}$ and the last step was obtained using $\frac{d\alpha}{dt} = 0$, as we assumed from the beginning. Denoting $\bm{\lambda}_i = \bm{\lambda}(t_i)$, $i = 1$ or $2$, we get
\[
\Delta E = \int_{\bm{\lambda}_1}^{\bm{\lambda}_2} \bm{\nabla}_{\lambda} g(\alpha,\bm{\lambda}) \cdot d\bm{\lambda} = g(\alpha,\bm{\lambda}_2) - g(\alpha,\bm{\lambda}_1),
\]
and the energy difference only depends on how $g$ changes through the variation of $\bm{\lambda}(t)$.

In our specific case, $\alpha$ is the angular momentum $L$, $H_0$ is the QS Hamiltonian with $a = 1$, $\bm{\lambda} = b$ and $g(\alpha,\bm{\lambda}) = b L$.
If the energy of a trajectory at the beginning of step $2$ is $E_2$ and at the end is $E_3$, their difference is
\begin{equation}\label{eq:e3}
E_3 = E_2 - |L|b_M,
\end{equation}
and the energy of every trajectory always decreases after the second step, independent on the switching time $\tau$.
It should now be clear why we set up our mechanical cycle like we did: using the information acquired in the measurement of angular momentum, we can make sure that all trajectories lose energy during the second step, and the bigger the angular momentum, the bigger the energy loss.

\subsubsection{Third step:}

During the third step of the mechanical cycle, we again make use of the Ergodic Adiabatic Theorem.
Denote by $E_3$ the energy of a trajectory at the beginning of the third step and by $E_4$ the energy of this same trajectory at the end of this same step.
Then
\[
\Omega(E_3,1,\pm b_M) = \Omega(E_4,0.1,0).
\]
Using \ref{eq:Omega_a,eq:Omega_b} and solving for $E_4$ gives
\begin{equation}\label{eq:e4}
E_4 = \left( \frac{\pi}{16f(0.1)} \left[ \left( b_M^4 + 4E_3 \right)^{3/2} + b_M^2 \left( b_M^4 + 6E_3 \right) \right] \right)^{2/3}.
\end{equation}
The mechanical cycle is finished.

\subsection{Energy variation in the entire mechanical cycle}\label{subsec:entire_cycle}

We can write the total energy variation $\Delta E = E_4 - E_1$ in the mechanical cycle as a function of the initial energy $E_1$, the angular momentum $L$ measured in the second step and $b_M$ (from \ref{eq:e2,eq:e3,eq:e4})
\begin{equation}\label{eq:delta_E}
\eqalign{
\fl \Delta E(E_1,L,b_M) = \left\{ \frac{\pi}{16f(0.1)} \bm{\left[} \bm{\left(} b_M^4 + 4 \left[ \left( \frac{2f(0.1)}{\pi} \right)^{2/3} E_1 - |L|b_M \right] \bm{\right)}^{3/2} \right. \right. \cr
\left. \left. + b_M^2 \left( b_M^4 + 6 \left[ \left( \frac{2f(0.1)}{\pi} \right)^{2/3} E_1 - |L|b_M \right] \right) \bm{\right]} \right\}^{2/3} - E_1.
} 
\end{equation}

\begin{figure*}

\newcommand{\scl}{.25}

\centering

\subfloat[\label{fig:E0.1_bM1}]{\includegraphics[scale=\scl]{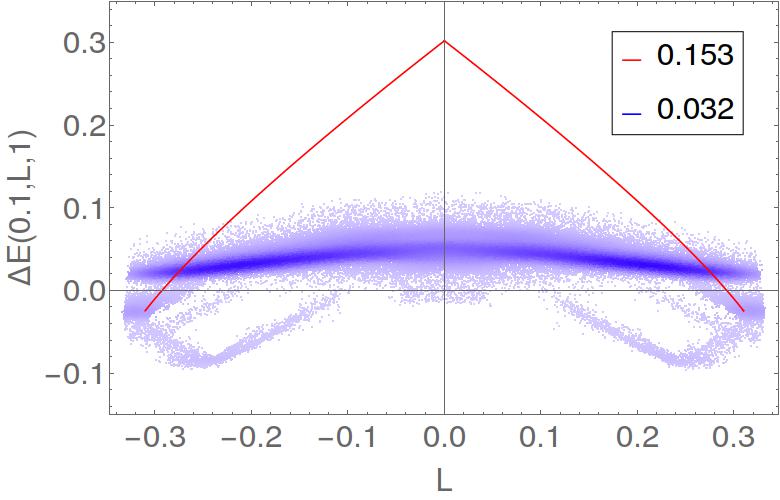}}
\qquad
\subfloat[\label{fig:E2_bM1}  ]{\includegraphics[scale=\scl]{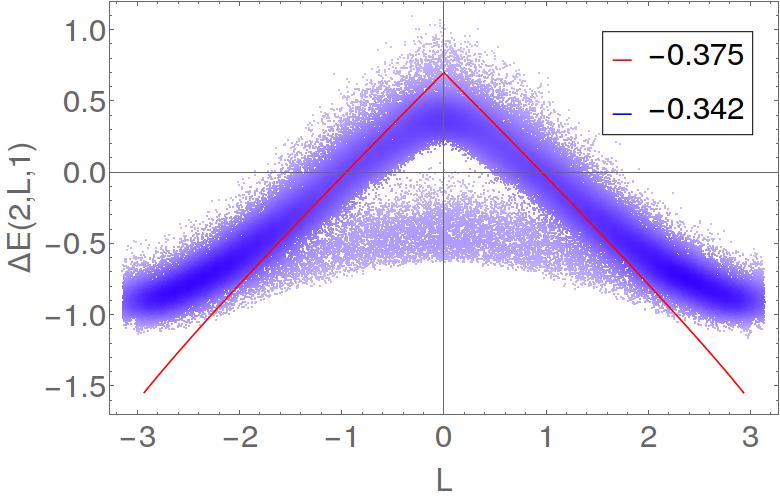}  }

\subfloat[\label{fig:E1_bM1}  ]{\includegraphics[scale=\scl]{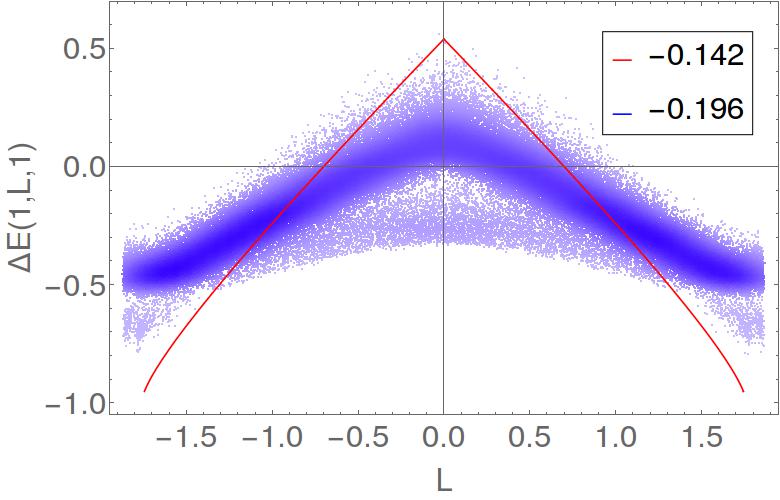}  }

\subfloat[\label{fig:E1_bM0.5}]{\includegraphics[scale=\scl]{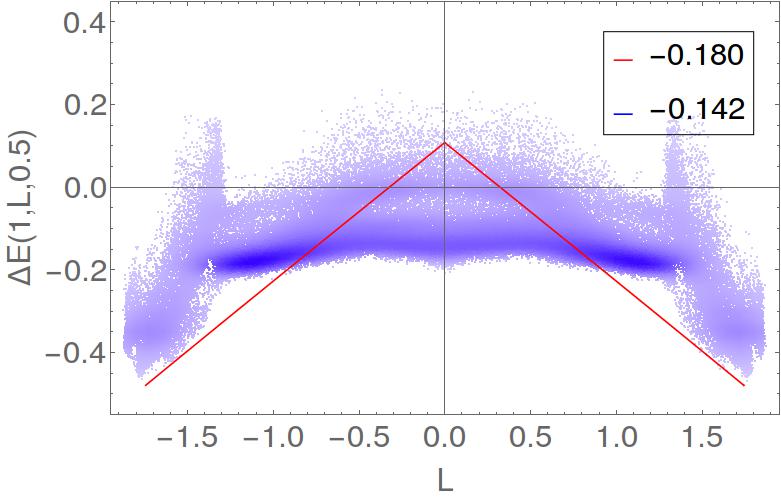}}
\qquad
\subfloat[\label{fig:E1_bM1.5}]{\includegraphics[scale=\scl]{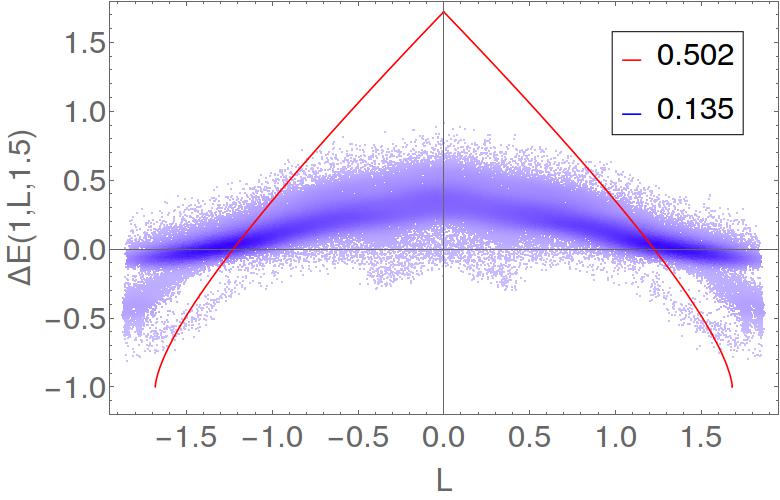}}

\caption{
(Color online) Comparison of the theoretical prediction $\Delta E(E_1,L,b_M)$, given by \ref{eq:delta_E}, and the data obtained from numerical simulations for a few values of $E_1$ and $b_M$.
In red, we have the energy variation versus angular momentum measured curve obtained by application the Ergodic Adiabatic Theorem and in blue we have the $10^5$ initial conditions sampled, with higher density of points represented by darker shades of blue.
The legends on each figure give the average energy variation, theoretical (red) and numerical (blue).
The parameters used in each numerical simulation are: (a) $E_1 = 0.1$, $b_M = 1$; (b) $E_1 = 2$,   $b_M = 1$;(c) $E_1 = 1$,   $b_M = 1$; (d) $E_1 = 1$,   $b_M = 0.5$; (e) $E_1 = 1$,   $b_M = 1.5$.
}

\label{fig:deXL_wf}

\end{figure*}

Even though this approach may be too simplistic, \ref{eq:delta_E} can be compared to numerical simulations of the system.
Using a sympletic integrator \cite{forest}, we sampled $10^5$ initial conditions for a few values of $E_1$ and evolved them through our mechanical cycle for a few values of $b_M$ in $\tau = 10^3 \tau_R$ units of simulation time.
\Cref{fig:deXL_wf} shows, for each numerical simulation with $E_1$ and $b_M$ given, the energy variation versus angular momentum measured of each initial condition, along with the theoretical prediction from \ref{eq:delta_E}.
The sign of the theoretical average energy variations can be estimated from the area under the red curves in \ref{fig:deXL_wf}.
This is so because the angular momentum distributions after the second step of the mechanical cycle do not show significant variation within the interval of allowed values of angular momentum and can be considered approximately uniform (see \ref{sec:angular_momentum_distribution}).
Hence, theoretical energy extraction can be easily visualized.
We can see that, in \ref{fig:E0.1_bM1,fig:E1_bM0.5,fig:E1_bM1.5}, the agreement between theory and data is far-fetched and the arrangement of points is too complicated to be well described by \ref{eq:delta_E}.
However, in \ref{fig:E1_bM1} and even more in \ref{fig:E2_bM1}, although there is no complete agreement between the simulation data and the theoretical prediction, the majority of points lie around a curve that roughly follows the theoretical curve.
These points present positive energy variation for low values of angular momentum and negative energy variation for high values of angular momentum, conforming to what we predicted the second step of the mechanical cycle would do.
There are also points that do not resemble the theoretical curve at all, forming a secondary curve that lies entirely below the horizontal axis $\Delta E = 0$, giving us initial conditions that lose energy during the mechanical cycle and that did not fit into our predictions.
\Cref{fig:E2_bM1} is also the one where the theoretical prediction of the average energy variation, $\langle \Delta E \rangle^{\mbox{\scriptsize the}}$ (see \ref{sec:angular_momentum_distribution} to understand how $\langle \Delta E \rangle^{\mbox{\scriptsize the}}$ is calculated), shows the least amount of percentual error compared to the numerical average, $\langle \Delta E \rangle^{\mbox{\scriptsize num}}$ (obtained by simple arithmetic average of the $10^5$ initial conditions), an error of approximately $9\%$.
On the other hand, \ref{fig:E0.1_bM1} shows the biggest error, approximately $375\%$.

Interestingly, data concurs with theory for all five numerical simulations in one aspect: the sign of the average energy variation.
For two of the simulations, \ref{fig:E0.1_bM1,fig:E1_bM1.5}, the average energy variation is positive, whereas, in the other three, it is negative.
This shows that energy extraction is not possible for any value of $E_1$ and $b_M$.
Nevertheless, for a certain value of $E_1$, there always seems to exist more than one value of $b_M$ that ensures negative energy variation.
The disagreement between data and theory can be attributed to the non-ergodicity of our system during the majority of the mechanical cycle and to the possibility that our mechanical cycle might never be able to be implemented adiabatically due to the presence of phase-space separatrices.
Not surprisingly, these two reasons are related to the two conditions necessary for the Ergodic Adiabatic Theory to hold true.
It is indeed unexpected that there is any approximate agreement between our theoretical predictions and the numerical data.

\begin{figure*}

\centering

\includegraphics[scale=.75]{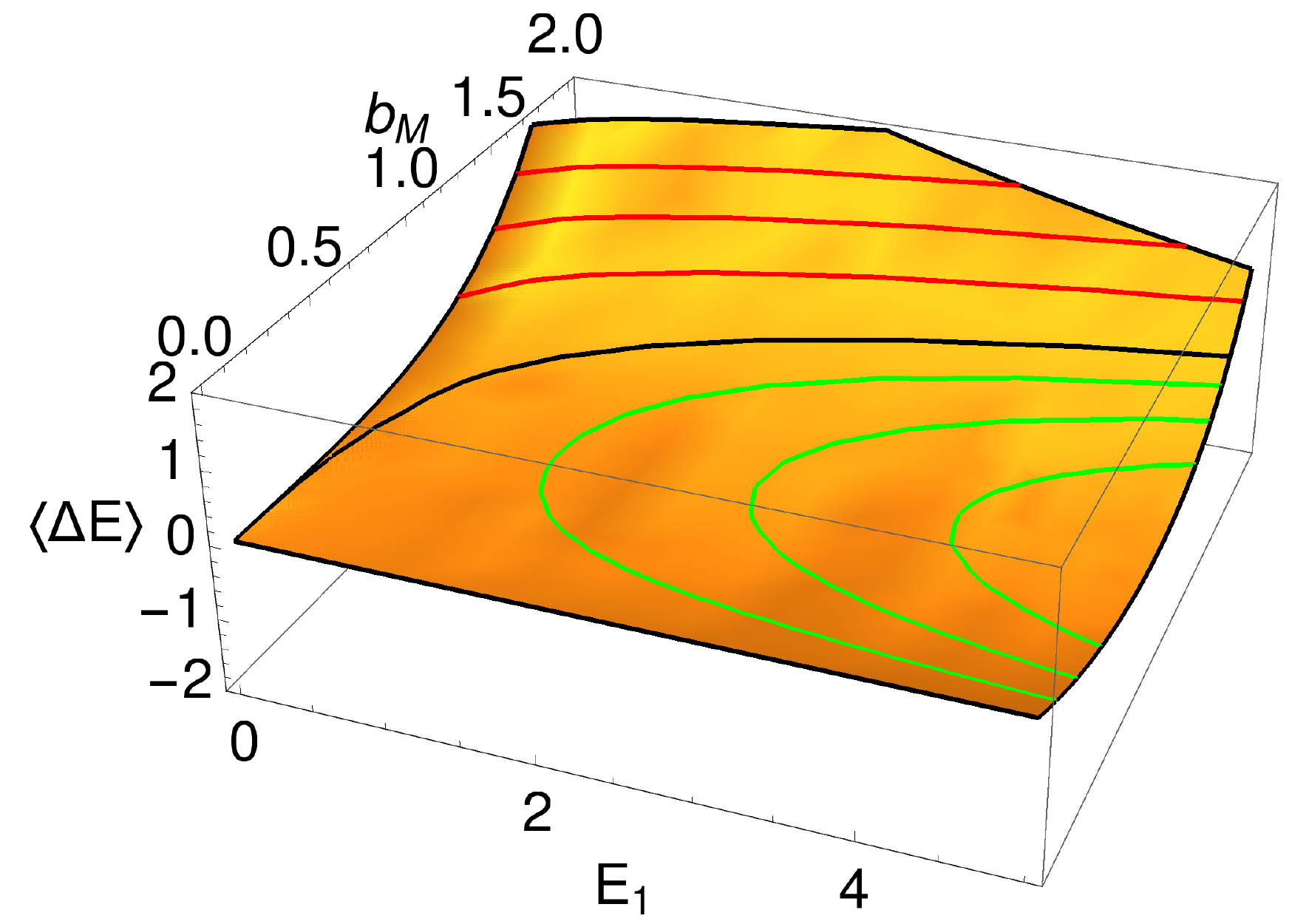}

\caption{
(Color online) 3D plot of the average expected energy variation from \ref{eq:delta_E_the}.
The green lines are contour lines of negative energy variation, the red lines are contour lines of positive energy variation and the black line corresponds to no energy variation.
The contour line of zero average energy variation is monotonic, which implies that, given $E_1$, there is always a value of $b_M$ that gives negative energy variation.
}

\label{fig:delta_E_the}

\end{figure*}

From \ref{eq:delta_E}, we can determine the average energy variation in the mechanical cycle, given $E_1$ and $b_M$,
\begin{equation}\label{eq:delta_E_the}
\langle \Delta E \rangle^{\mbox{\scriptsize the}} (E_1,b_M) = \int \Delta E (E_1, L, b_M) \rho(L|E_2) dL,
\end{equation}
where $\rho(L|E_2)$ is the angular momentum distribution for possible values of $L$ of a trajectory after step 1 of the mechanical cycle, conditioned by that trajectory having energy $E_2$, given by \ref{eq:rho_L}.
\Cref{fig:delta_E_the} shows a 3D plot of \ref{eq:delta_E_the}, together with a few contour lines.
It shows that, no matter the value $E_1$ in the beginning of the mechanical cycle, there exists a value of $b_M$ that ensures negative average energy variation.

The numerical simulations confirms that the microcanonical average of $\Delta E$ is smaller than zero for the specific values of initial energy $E_1$ and external parameter $b_M$ we used.
Even if that is not the case for all values of $E_1$ and $b_M$, the mere existence of a such an example defies the Kelvin-Planck statement of the Second Law of Thermodynamics, as argued in the Introduction.

\section{Discussion}\label{sec:discussion}

Our model is very similar to the Microcanonical Szil\'ard Engines existent so far.
With the same general setup as ours (starting with a system in contact with a heat reservoir, disconecting them, acting on the system and finally reconnecting them), Vaikuntanathan and Jarzynski \cite{modeling} devised a one-dimensional system in which they can consistently lower the system's energy in a feedback process with energy measurement, no matter the energy measured.
The critical difference here is that our energy variation in only negative \it{on average}, while theirs is negative in every single realization of their process (at least in the quasi-static limit, some trajectories do gain energy for finite times).
They too apply the Ergodic Adiabatic Theorem to show the extraction of energy and, in another instance \cite{apparent}, show that the presence of separatrices in phase-space during their process is crucial for their engine to work.
Because of this, we believe that phase-space separatrices are relevant in our case too, but a more meticulous investigation would be required to understand the effects of separatrix crossings here, either in enabling negative average variation or in justifying the disagreement between data and theory in \ref{fig:deXL_wf}.
One thing to note is that separatrices in our model ought to be much more complicated than those found in Vaikunthanathan and Jarzynski's model, where a closed form for their equation can be found effortlessly.
Studies of the effects of separatrix crossing in one-dimensional systems have been developed \cite{separatrix1,separatrix2}, but not much has been done to multi-dimensional systems.

Our model has a lot in common with the original Szil\'ard Engine too.
When the barrier is introduced in the Szil\'ard Engine, the particle ``chooses'' a side of the box and the phase space reflection symmetry with respect to the coordinate axis (a parity symmetry) is broken.
Half of the phase space is essentially erased for that particular trajectory after the measurement.	
With the assistance of measurement, a protocol is carried out in order to abuse this lack of reflection symmetry and extract energy from the system.
This may seem at odds with our model, because right before the measurement is made, when $a = 1$, a symmetry (angular momentum conservation) is established, not broken.
Nonetheless, what matters here is the reflection symmetry that is broken when the particle ``chooses'' a certain value of angular momentum.
After the measurement, step 2 of the mechanical cycle makes sure the energy of that specific trajectory decreases.
In a way, we are using the two regions of positive and negative angular momentum of \ref{fig:poincare_sections_1.0} as the two sides of the boxes of a Szil\'ard Engine, and it is precisely the conservation of angular momentum that allows us to brake the reflection symmetry.
A key difference between our model and the original Szil\'ard Engine is that, in the original Szil\'ard Engine, the barrier insertion and removal have no energy costs, while the equivalent in our model (steps 1 and 3 of the evolution of external parameters, respectively) do come with energy variations.

It has been noted in reference \cite{phase} that examples of Szil\'ard engines, all of which include feedback processes, usually apply processes that violate either Liouville's Theorem (like the original Szil\'ard Engine \cite{szilard}) or the Ergodic Adiabatic Theorem (like the Microcanonical Szil\'ard Engine of reference \cite{modeling}).
Considering that these are the two theorems used to show the impossibility of energy extraction from a single heat reservoir in a cycle, it is no surprise that Szil\'ard engines violate the Kelvin-Planck statement of the second law.
As mentioned in the Introduction, it has been shown, in a very general context, that the mistreatment of Liouville's theorem in isothermal processes is linked to symmetry breaking \cite{universal,revisited}, in the sense that the phase-space loses a symmetry that the Hamiltonian has.
Such an analysis has not yet been developed for thermally isolated processes, that happen without connection to a heat reservoir, like the processes carried out in Microcanonical Szil\'ard Engines.

However, there is a case to be made that separatrix crossing, present in Microcanonical Szil\'ard Engines, constitutes a kind of symmetry breaking.
After all, separatrix crossing is a possible cause of sudden shrinkage of phase-space volume.
Separatrix crossing symmetry breaking is not a symmetry breaking caused directly by an external agent like in the original Szil\'ard Engine, where the insertion of a barrier causes sudden shrinkage of phase space volume, but a symmetry breaking caused by the natural evolution of a time-dependent system.
The external agent still acts indirectly, creating and destroying separatrices through the variation of external parameters and ultimately shrinking phase-space volume whenever a trajectory crosses a separatrix.
In the end, whenever an isolated system (and that may be system of interest plus heat reservoir) suddenly experiences a decrease in its phase-space volume $\Omega$, as caused by any type of symmetry breaking, the system's entropy $ S \propto \log \Omega $ will also decrease and it is easy to see how the Second Law of Thermodynamics fails, as per Planck's statement.

\section{Conclusions}\label{sec:conclusion}

In this paper, we described how to build a system with two degrees of freedom that goes against the Kelvin-Planck statement of the Second Law of Thermodynamics.
We used a theoretical tool, the Ergodic Adiabatic Theorem, and numerical simulations to motivate and show exactly how our model works.
Our model is, by all means, a Maxwell Demon.

Reclaiming the Kelvin-Planck statement from the demon via Landauer's principle in our case is done in the same way Vaikuntanathan and Jarzyski \cite{modeling} have done for their case.
We can consider a finite precision energy measurement apparatus that saves the information acquired in each measurement in a certain number of bits.
We then choose the value of our external parameter, $b_M$, based on this finite precision measurement, as with any feedback process.
We can see, from \ref{fig:delta_E_the}, that it is always possible to choose a value of $b_M$ that guarantees average energy extraction on average.
Applying here the same reasoning given in section III of reference \cite{modeling}, which includes Landauer's principle, it follows that the maximum work extracted with finite precision measurements from the system only matches the minimum work required to erase the information contained in the bits.
Even more energy would be necessary to erase the angular momentum information saved, as that would require one extra bit (the sign of the angular momentum can be positive or negative) per energy measurement.

Our results contrast with the results obtained by the authors of reference \cite{robust}, regarding the \it{Fermi acceleration} phenomenon.
They show that a particle in a two-dimensional time-dependent billiard, where the particle is constantly colliding with moving walls, has an exponential energy growth.
The billiard oscillates cyclically and adiabatically, just as our system (and most Microcanonical Szil\'ard Engines) does.
As we see it, there are two main differences: the absence of measurement in their system, and it is a well known fact that Szil\'ard Engines and Maxwell Demons all require information (i.e. feedback processes) to achieve their goal; and the absence of thermalization in their system, achieved in our system through the reconnection with the heat reservoir at the end of the thermodynamic cycle.
It remains to be shown exactly how these differences bring forth such differing outcomes.

Whether or not the discussion of the previous section is enough to qualify our example (and Sato's example) as a Microcanonical Szil\'ard Engine is open to debate.
Either way, in a sense, our example serves as a complement to the already established Microcanonical Szil\'ard Engines, showing that it is possible to conceive a system with more than one degree of freedom whose microcanonical average energy decreases after a cycle.
The fact that such a system can exist even when non-integrability and chaos are present suggests that Microcanonical Szil\'ard Engines are not limited to microscopic systems and that there should exist a way to explain Microcanonical Szil\'ard Engines through symmetry breaking.

\appendix

\section{Averages in the microcanonical ensemble}\label{sec:averages_microcanonical_ensemble}

In this appendix, we show how to calculate microcanonical averages fo the QS Hamiltonian given by \ref{eq:QS}
\begin{equation}\label{eq:QS_appendix}
H^{\mbox{\scriptsize QS}}(\bm{z},a) = \frac{p_x^2}{2} + \frac{p_y^2}{2} + \frac{a}{4}\left( x^4 + y^4 \right) + \frac{x^2 y^2}{2}.
\end{equation}
The microcanonical phase space distribution is
\begin{equation}\label{eq:microcanonical_distribution_z}
\rho(\bm{z},E,a) = \frac{\delta\left( H^{\mbox{\scriptsize QS}}(\bm{z},a) - E \right)}{\int \delta\left( H^{\mbox{\scriptsize QS}}(\bm{z},a) - E \right) d\bm{z}},
\end{equation}
where $\delta$ represents the Dirac delta function.
For easier manipulation of this distribution, we define a non-canonical transformation from the canonical variables $(x,p_x,y,p_y)$ to the non-canonical variables $(H,\psi,\theta,\phi)$:
\begin{equation}\label{eq:variable_transformation}
\begin{array}{ccl}
x^2 & = & \sqrt{\frac{2H}{\cos 2\theta}} \left( \frac{\cos \theta}{\sqrt{1+a}} + \frac{\sin \theta}{\sqrt{1-a}} \right) \sin\psi;\\
y^2 & = & \sqrt{\frac{2H}{\cos 2\theta}} \left( \frac{\cos \theta}{\sqrt{1+a}} - \frac{\sin \theta}{\sqrt{1-a}} \right) \sin\psi;\\
p_x & = & \sqrt{2H} \cos\phi \cos\psi;\\
p_y & = & \sqrt{2H} \sin\phi \cos\psi,
\end{array}
\end{equation}
with $(H,\psi,\theta,\phi)$ well defined within $0 \leq H \leq \infty, 0 \leq \psi \leq \pi/2, 0 \leq \theta \leq \cos^{-1}(a)/2$ and $0 \leq \phi \leq 2\pi$. 
This way, the kinetic and potential parts of \ref{eq:QS_appendix} can be written as
\begin{equation}\label{eq:kinetic_energy}
K = \frac{p_x^2}{2} + \frac{p_y^2}{2} = H \cos^2 \psi
\end{equation}
and
\begin{equation}\label{eq:potential_energy}
V = \frac{a}{4}\left( x^4 + y^4 \right) + \frac{x^2 y^2}{2} = H \sin^2 \psi,
\end{equation}
while \ref{eq:microcanonical_distribution_z} simplifies to
\begin{equation}\label{eq:microcanonical_distribution_H}
\rho(H,E,a) = \frac{\delta( H - E )}{\int J(E,\psi,\theta,\psi) d\psi d\theta d\phi},
\end{equation}
where
\[
J(H,\psi,\theta,\phi) = \sqrt{\frac{H}{2(\cos 2\theta - a)\cos 2\theta}} \cos\psi
\]
is the Jacobian of the non-canonical transformation defined by \ref{eq:variable_transformation}.
The average kinetic energy is then, with help from \ref{eq:kinetic_energy,eq:microcanonical_distribution_H},
\[
\eqalign{
\fl \langle K \rangle(E,a) = \int K  \rho(\bm{z},E,a) d\bm{z} \cr
\fl \langle K \rangle(E,a) = \frac{1}{\int J(E,\psi,\theta,\psi) d\psi d\theta d\phi} \int H \cos^2\psi\,\delta( H - E ) J(H,\psi,\theta,\phi) dH d\psi d\theta d\phi \cr
\fl \langle K \rangle(E,a) = E \int_0^{\pi/2} \cos^3\psi d\psi = \frac{2E}{3}.
}
\]
Likewise, from \ref{eq:potential_energy,eq:microcanonical_distribution_H}, the average potential energy is
\[
\langle V \rangle(E,a) = E \int_0^{\pi/2} \sin^2\psi \cos\psi d\psi = \frac{E}{3}.
\]
Non surprisingly, we have $\langle K \rangle + \langle V \rangle = E$.

\section{Angular momentum distribution}\label{sec:angular_momentum_distribution}

In this appendix, we will show the angular momentum distribution for the Hamiltonian of \ref{eq:QS} and how it is used to calculate theoretical average energy variations in our mechanical cycle.

Assuming the Ergodic Adiabatic Theorem holds true, every trajectory that starts the mechanical cycle with energy $E_1$ will reach the end of the first step of the mechanical cycle with energy $E_2$, given by \ref{eq:e2}.
The angular momentum distribution $\rho(L|E_2)$ for possible values of angular momentum $L$ of the trajectories, conditioned by having energy $E_2$ at the end of the first step, is
\begin{equation}\label{eq:rho_L}
\rho(L|E_2) = \frac{1}{\omega(E_2,1)} \int \delta \left( E_2 - H^{\mbox{\scriptsize QS}}(\bm{z},1) \right)
\delta \left( L - (x p_y - y p_x) \right) dz,
\end{equation}
where $\omega(E,a) = \int \delta \left( E - H^{\mbox{\scriptsize QS}}(\bm{z},a) \right) dz$ is the density of states of energy $E$ and $\delta$ is the Dirac Delta function.
This distribution is nothing more than the summation of contributions of all trajectories with angular momentum $L$ weighted by the microcanonical distribution $\frac{1}{\omega(E,a)} \delta \left( E - H^{\mbox{\scriptsize QS}}(\bm{z},a) \right) $.
The average energy variation of the mechanical cycle is then given by \ref{eq:delta_E_the}.

\begin{figure*}

\centering

\subfloat[\label{fig:hist_L_e=0.5}]{\includegraphics[scale=.7]{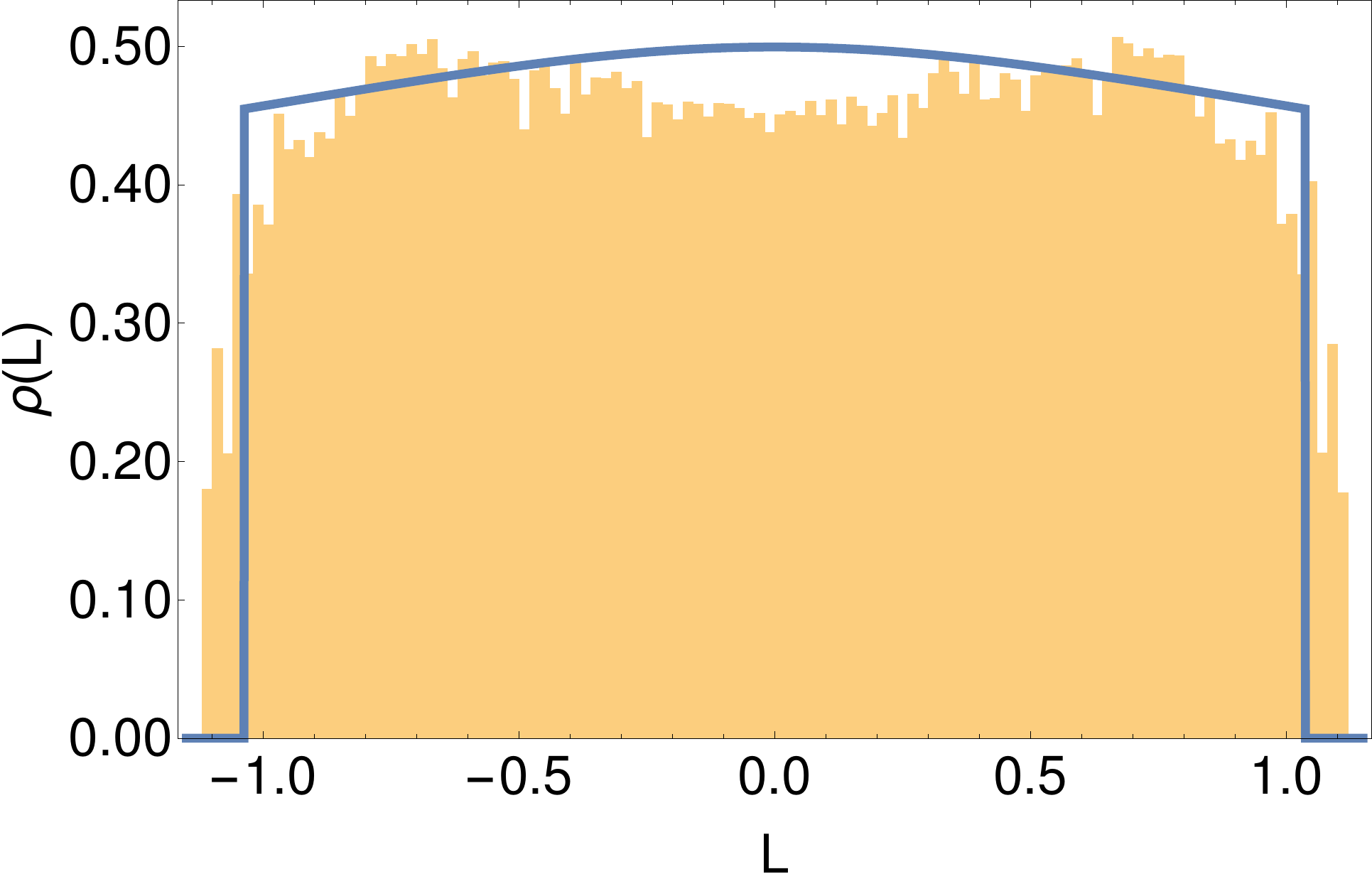}}

\subfloat[\label{fig:hist_L_e=2}]{\includegraphics[scale=.7]{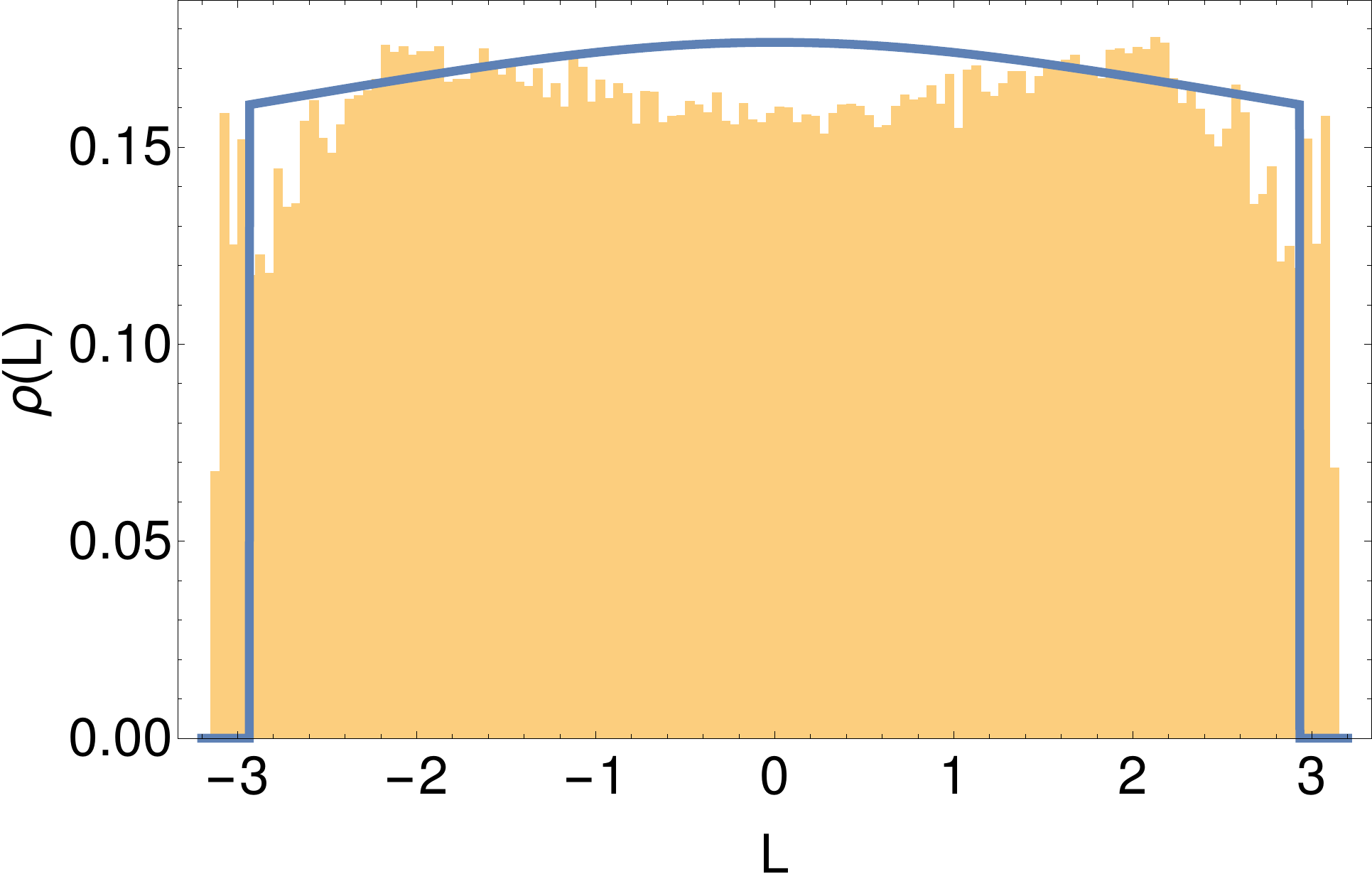}}

\caption{
(Color online) Histograms of angular momentum $L$ after the first step of the mechanical cycle for (a) $E_1=0.5$ and (b) $E_1=2$, where $\rho(L)$ is the angular momentum distribution.
The blue curves represent the expected angular momentum distribution from the Ergodic Adiabatic Theorem, \ref{eq:rho_L}.
The sudden drop to zero is expected: $L$ has a maximum and a minimum possible value as functions of the system's energy $E_2$, given by $L^{\mbox{\scriptsize max}} = \left( \frac{4E_2}{3} \right)^{3/4}$, and so this distribution should be zero for values outside the range $-L^{\mbox{\scriptsize max}} < L < L^{\mbox{\scriptsize max}}$.
}

\label{fig:hist_L}

\end{figure*}

\Cref{fig:hist_L} shows the distribution of \ref{eq:rho_L} for $E_1 = 0.5$ (\ref{fig:hist_L_e=0.5}) and $E_1=2$ (\ref{fig:hist_L_e=2}), together with histograms of angular momenta after the first step of the mechanical cycle described in \ref{sec:mechanical_cycle}, obtained through numerical simulations.
We can also see that the theoretical curves deviate a lot from the numerical simulations, which might be a sign of separatrix crossing invalidating the Ergodic Adiabatic Theorem.

\ack
The authors acknowledge support from FAPESP/CAPES, grant \#2016/19631-9, S\~ao Paulo Research Foundation (FAPESP).


\section*{References}

\bibliographystyle{plain}

\end{document}